\documentclass{aa}

\usepackage{graphicx}
\usepackage[colorlinks=true,linkcolor=blue,citecolor=blue]{hyperref}
\usepackage{txfonts}

\begin{document}

   \title{Magnetically--gated accretion model: application to short bursts in the intermediate polar V1223 Sgr }

   \author{J.-M. Hameury
          \inst{1}
          \and
          J.-P. Lasota\inst{2,3}
          \and
          A. W. Shaw\inst{4}
          }

   \institute{Observatoire Astronomique de Strasbourg, CNRS UMR 7550, 67000 Strasbourg, France\\
             \email{jean-marie.hameury@astro.unistra.fr}
         \and
             Institut d'Astrophysique de Paris, UMR 7095 CNRS, UPMC Universit\'e Paris 6, 98bis Bd Arago, 75014 Paris, France
         \and
             Nicolaus Copernicus Astronomical Center, Polish Academy of Sciences, Bartycka 18, 00-716 Warsaw, Poland
         \and
             Department of Physics, University of Nevada, Reno, NV 89557, USA
             }

   \date{Received / Accepted}


  \abstract
   {Some intermediate polars show outbursts that are much shorter than those observed in normal dwarf novae, and whose origin has remained unelucidated for a long time. }
   {We examine here the case of V1223 Sgr, an intermediate polar that showed a short outburst in 1984, and compare the outburst characteristics with the predictions of the magnetospheric gating model.}
   {We use the archival data from the AAVSO from which we extract the outburst profiles. We use our code for computing the time-dependent evolution of an accretion disc truncated by the white dwarf magnetic field, using a simple description of the interaction between the disc and the magnetic field, as in \citet{ds10}.}
   {We find that V1223 Sgr underwent a series of short outbursts, with a rise lasting for typically two to three hours, and a slightly longer decay. When applied to intermediate polars, the model by \citet{ds10} accounts well for the observed outburst duration and intensity. We confirm, however, that the model outcome depends sensitively on the rather poorly constrained model's assumptions. We have also searched the AAVSO database for short outbursts in other IPs, identifying individual short outbursts in FO Aqr, TV Col, NY Lup and EI UMa, but no series as those observed in V1223 Sgr. We also found a superoutburst, followed by a reflare in CTCV J2056-3014.}
   {Although the magnetic--gating accretion instability model is clearly responsible for the series of V1223 Sgr short outbursts and most probably for similar events in other intermediate polars, the model describing this process needs improving, in particular concerning the interaction of the white--dwarf's magnetic field with the accretion disc. This difficult task might benefit from further comparison of the model outcome with additional observations having good time coverage and time resolution.}

   \keywords{accretion, accretion discs -- stars: dwarf novae -- instabilities
               }

   \maketitle
%

\section{Introduction}
Dwarf nova (DN) eruptions are sudden brightenings by typically a few magnitudes that are observed in some classes of cataclysmic variables. They usually last for a few days, and recur on time scales of weeks \citep[see e.g.][for a detailed review of these objects]{w03}. It has been known since long and it is now very clear \citep{dol18} that these outbursts are due to a thermal-viscous instability of the accretion disc \citep[see][for reviews]{l01,h20}. The instability occurs when, somewhere in the disc, the mid-plane temperature is in the range 10,000 -- 20,000~K, such that hydrogen is partially ionized and the opacities strongly dependent on temperature. During low states, the disc is cold and neutral everywhere while the mass accretion rate onto the white dwarf is small; the disc mass builds up and the central temperature increases,eventually reaching the critical value at which ionization becomes important and the instability sets is. Heat fronts propagate bringing the accretion disc into a hot state in which the accretion rate is higher than the rate at which the secondary transfers mass to the disc. The disc then empties and cools down, until a point where it cannot be maintained in a hot state; a cooling wave starts from the disc outer edge, crosses the disc bringing it to quiescence.

Intermediate polars (IPs) are cataclysmic variables in which the white dwarf magnetic field is strong enough to prevent the accretion disc to reach the white dwarf surface, or even to fully prevent its formation, but not strong enough to force corotation of the white dwarf with the orbit. If a disc exists, the disc instability model (DIM) can still apply, and IPs can also exhibit dwarf nova outbursts, as discussed in \citet{hl17}; typical dwarf nova outbursts have been observed in IPs such as, for example, GK Per. Some IPs, however, show short outbursts (a few hours) that cannot be explained by the DIM. Until recently, the best documented example was TV Col \citep{sm84,hb93,hss05} which shows outbursts lasting for about six hours. Short outbursts have also been detected in IPs such as V1223 Sgr \citep{vv89} or CXOGBS J174954.5-294335 \citep{jth17}, but due to insufficient time coverage, only upper limits on the outburst duration could be obtained. The situation changed  with the observation by TESS of V1025 Cen, which showed a rapid succession (recurrence time ranging between one and 3 days) of short (less than six hours) outbursts \citep{llh22}. \citet{csk22} also reported the detection of many short outbursts in V1223 Sgr that we explore in detail here. Cataclysmic variables not classified as magnetic have also shown short outbursts; MV Lyr experienced a series of 30 minutes outbursts every two hours \citep{smd17}. TW Pic \citep{sdb22} had abrupt drops by factors up to 3.5 on timescales as short as 30 minutes; in the low mode, it displayed bursts with a recurrence time of 1.2 -- 2.4~h. In both cases, a weak magnetic field was invoked to account for the observations.

Despite early claims that the truncation of the accretion disc by the white dwarf magnetic field results in shorter bursts than in non magnetic CVs, and that outbursts lasting for a few hours in CVs could be explained by the DIM \citep{svv88}, it was soon realized that this cannot be the case. Using the DIM, \citet{av89} did not get outbursts lasting less than one day, even though they used a viscosity parameter $\alpha$ in the hot state as high as unity. It has also been suggested that these short outbursts are due to a mass transfer instability occurring in the secondary  \citep[see, e.g.][]{av89,hb93}, but calculations by \citet{hl17} using reasonable values for the viscosity parameter in the disc showed that this would require the disc to remain in the cold state and necessitate large changes in the mass transfer rate in order to avoid light fluctuations on the viscous time scale, longer than the outburst duration. The outburst would then solely be due to an increase of the bright spot luminosity.

It has been recently proposed that short outbursts are ``micronovae" due to localized thermonuclear explosions at the surface of the white dwarf, as a result of confinement of the accreted material by the magnetic field \citep{sgk22a,sgk22b}. This, however, relies on the hypothesis that the magnetic field is rigidly anchored at the white dwarf surface; \citet{hl85} showed that, if this hypothesis holds, the magnetic field can indeed confine the accreted material even if the magnetic pressure is much less than the gas pressure; they noted, however, that, in contrast with the neutron star case, this is rather uncertain. In any case, whereas such a model could account for isolated outbursts observed in e.g. TV Col, it cannot explain sequences of outbursts recurring on timescale of days or less \citep{sgk22a}.

It was finally proposed by \citet{llh22} that the rapid succession of short outbursts in V1025 Cen  are due to magnetospheric gating, initially proposed by \citet{st93} to explain the rapid burster phenomenon and later developed by \citet{ds10,ds11,ds12}, hereafter DS10, DS11 and DS12 respectively. This followed  an earlier suggestions by \citet{mbd07} that EX Hya outbursts are due to this mechanism, and the proposals by \citet{smd17,sdb22} that this mechanism is also at play in MV Lyr and TW Pic provided that the white dwarf in these two objects is weakly magnetized, with a magnetic field in the range $2 \times 10^4$ -- $10^5$~G for MV Lyr and less than $10^6$~G in TW Pic.

In Sect. \ref{v1223} of this paper, we present in detail the characteristics of the outbursts observed in 2020 in V1223 Sgr. In Sect. \ref{sec:short} we present results of a search for short outbursts in other IPs from observations in the database maintained by the American Association of Variable Star Observers (AAVSO). We adapt the magnetospheric gating model to the IPs in Sect. \ref{model}, using it for the first time to calculate the light curves of short outbursts observed in these systems, applying it in particular to the case of  V1223~Sgr. Our conclusions are summarized in Sec. \ref{sec:concl}

   \begin{figure}
   \centering
   \includegraphics[width=\columnwidth]{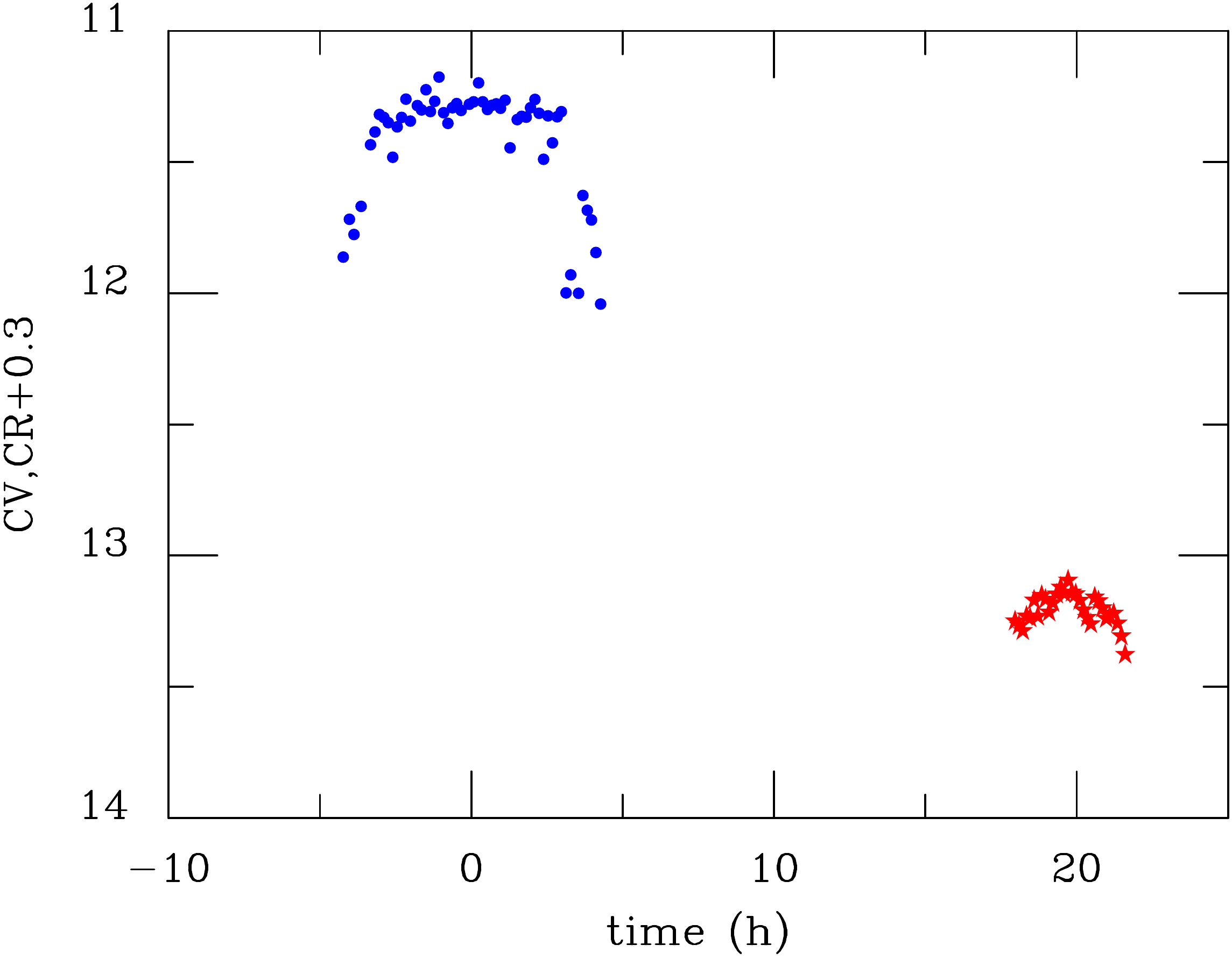}
   \caption{Time profile of the V1223 Sgr 2014 outburst (data from AAVSO). The visual $CV$ and red $CR+0.3$ data are shown by blue dots and red stars respectively.}
   \label{out2014}%
   \end{figure}

   \begin{figure*}
   \centering
   \includegraphics[width=\textwidth]{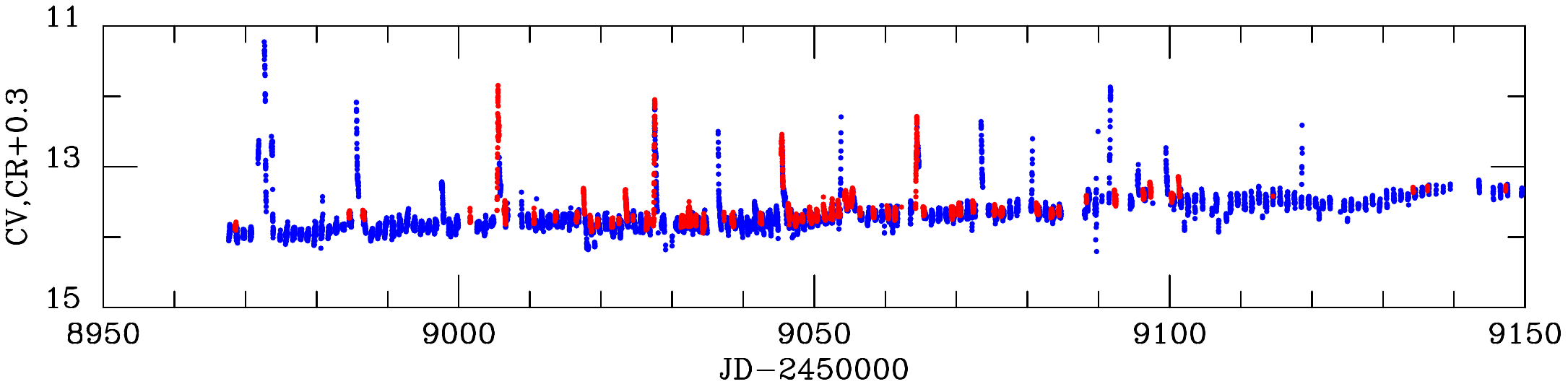}
   \caption{Light curve of V1223 Sgr (data from AAVSO). The unfiltered visual magnitude with Bessel $V$ zeropoint, $CV$, is shown in blue; the red points show the red magnitude $CR$ plus 0.3.}
   \label{lc}%
   \end{figure*}
   
\section{Rapid outbursts in V1223 Sgr} \label{v1223}

V1223 Sgr is an intermediate polar with an orbital period of 3.36~h and a spin period of 0.207~h. The primary and secondary masses, as determined from spectral fitting of the X-ray spectrum are $M_1=0.92$~M$_\odot$ and $M_2=0.33$~M$_\odot$ respectively \citep{hk21}, implying an inclination of $53\pm 2^\circ$; \citet{hk21} note that if the inner disc is truncated at the corotation radius, the primary mass should be revised to $M_1=0.97$~M$_\odot$, significantly larger than a previous determination by \citet{p85} who found that $M_1$ should be in the range 0.4 -- 0.6~M$_\odot$ and by \citet{shm18} who found $M_1=0.75$~M$_\odot$, but in line with determinations by \citet{hi14} and by \citet{bhm04} who obtain $M1 = 0.87$ and 0.93~M$_\odot$ respectively. In the following, we assume $M_1 = 0.9$~M$_\odot$. The parallax of the source, as determined by {\it GAIA}, is $1.72 \pm 0.047$~mas, corresponding to a distance of 580~pc \citep{g18}.

The spin period of V1223 Sgr has steadily increased over the past forty years \cite{pdk20}, which comes as a surprise since this source is bright, and would be expected to spin up, as do most other bright IPs. This questions the assumption that V1223 Sgr is in spin equilibrium, spending equal amounts of time spinning up during high states, and down during very low states, as noted by \citet{pdk20}. This makes difficult estimating the  magnetic moment $\mu$ of the primary from the spin period but \citet{nws04} suggest $\mu = 4 \times 10^{32}$~G~cm$^3$ for this source.

The AAVSO light curve shows that V1223 Sgr is most often at a constant magnitude $V=13$. Its 1984 outburst \citep{vv89} was not detected by the AAVSO, but another short outburst, presumably isolated, was detected on June 16, 2014; it lasted for at least 10~h, and reached a magnitude $V=11.2$. Its profile is shown in Fig.~\ref{out2014}, where we plot the visual $CV$ and $CR+0.3$, $CR$ being the unfiltered red magnitude with Bessel $R$ zeropoint. We added 0.3 to $CR$; when simultaneous $CV$ and $CR$ data are available, $CV \sim CR+0.3$. The shape of the outburst is very peculiar: it shows a very sharp rise and decay, with a flat top, quite dissimilar to what has been observed in 1984 and also dissimilar to the short outbursts observed in other IPs.

\begin{figure*}
   \centering
   \includegraphics[width=\textwidth]{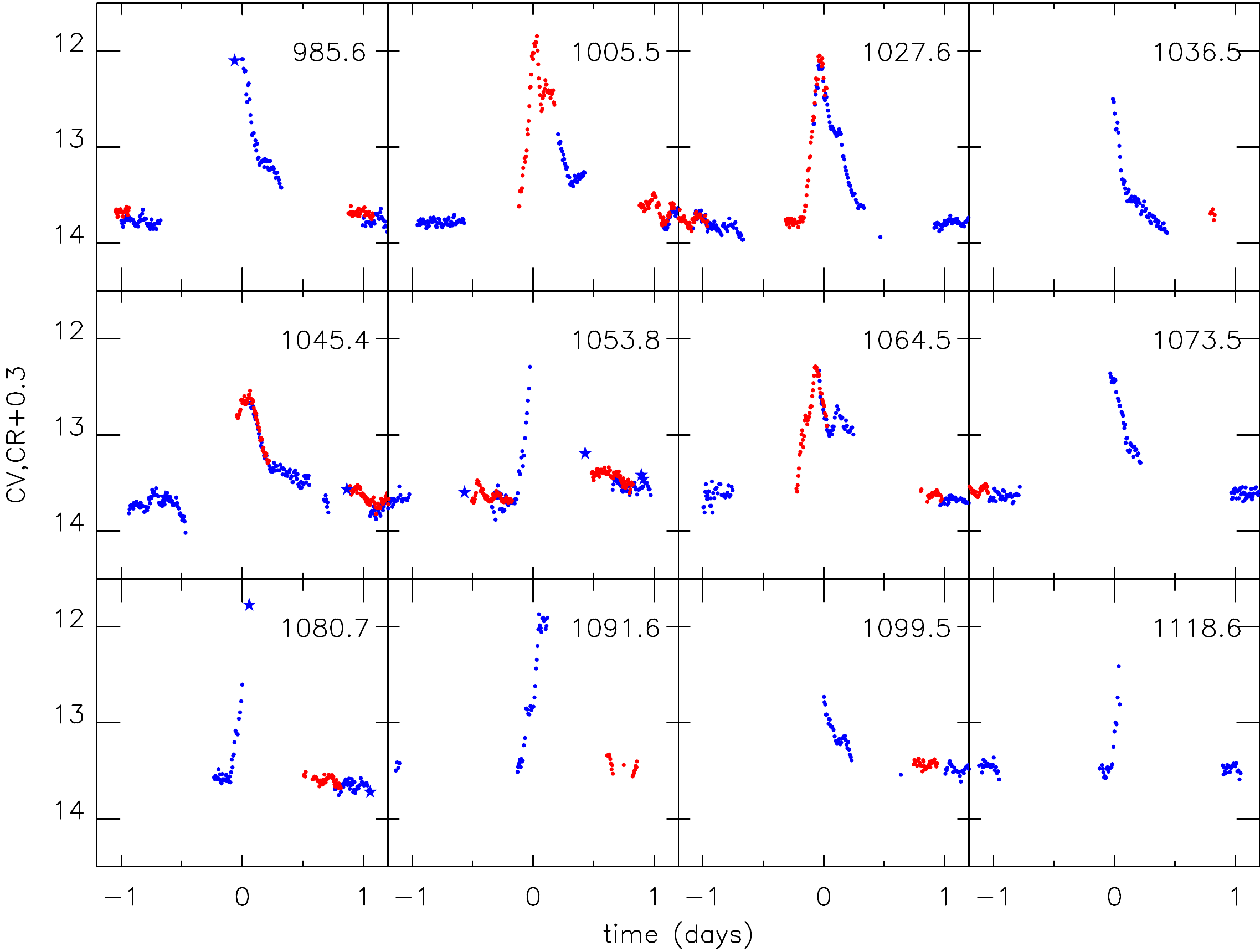}
   \caption{Details of the of V1223 Sgr outbursts (data from AAVSO). The stars represent data from the ASAS-SN. $CV$ data are shown in blue; the red points show the red magnitude plus 0.3. Time $t=0$ is ascribed to the peak of the observed flux; the label in the upper right corner of each frame is the Julian date of the outburst peak minus 2458000.}
   \label{outbursts_all}%
   \end{figure*}

V1223 Sgr went into a low state in 2018 -- 2019, as shown in \citet{csk22}. At the end of the normal--state recovery phase, the system underwent a series of outbursts that are shown in Fig.~\ref{lc}. The profiles of the twelve brightest outbursts are shown in Fig.~\ref{outbursts_all}, with the exception of the first outburst for which data obtained by two different observers at the same time and in the same band differ by as much as 1.3~mag. If the series of data points at $V\sim 13$ during the first outburst, as well as the peak of the outburst were overestimated by 1.3 mag., this would make the first outburst very similar to the other outbursts, both in duration and in peak luminosity, but this cannot be ascertained. One can note that the colour $CV-CR$ remains approximately equal to 0.3 both in quiescence and at the outburst peak. We have included in Fig. \ref{outbursts_all} data from the All-Sky Automated Survey for Supernovae \citep[ASAS-SN,][]{spg14,kss17}, which has not as good a sampling as the AAVSO data. Two outbursts have been detected both by the AAVSO and the ASAS-SN, at JD 2458985 and 2459080, with possibly a third one at JD 2459053.

Between JD 2458970 and JD 2459200, observational gaps longer than 1, 5 and 15~h represent 67\%, 64\% and 40\% respectively of this period, meaning that one probably missed almost as many bursts as observed. Given that 12 bursts reaching V=12.5 were detected over a 150 days interval, this would lead to an effective recurrence time of order of 6 days. The ASAS-SN observation windows of V1223 Sgr during the oubursting period do not differ much from those of the AAVSO, and do not bring additional constraints on missed outbursts.

\begin{figure}
   \centering
   \includegraphics[width=\columnwidth]{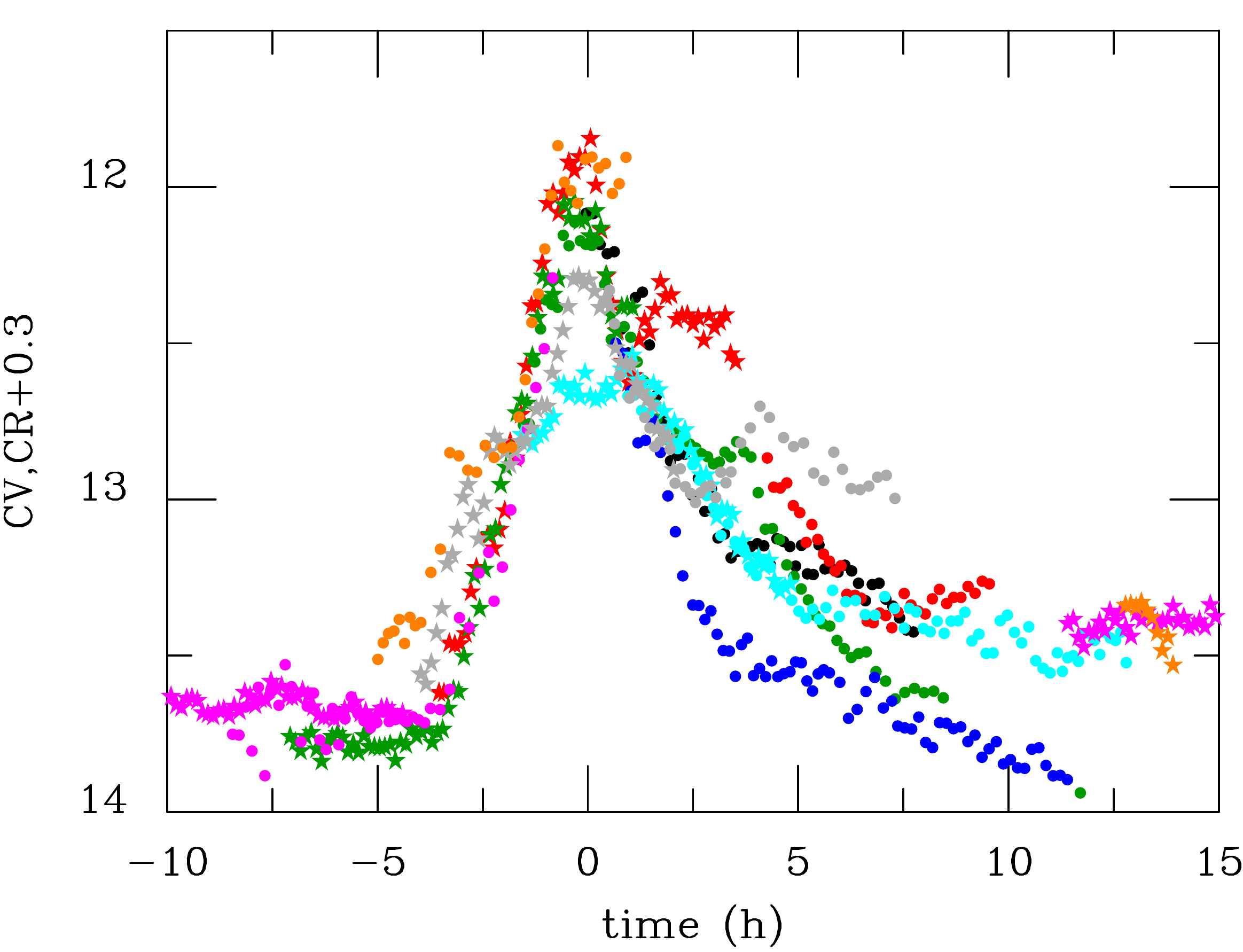}
   \caption{Light curve of V1223 Sgr (data from AAVSO). Each colour corresponds to a different outburst.}
   \label{outbursts}%
   \end{figure}

The outbursts profiles, as shown in Fig.~\ref{outbursts_all}, look quite similar. Figure~\ref{outbursts} shows in more detail the outburst profiles of the eight outbursts which have the best time coverage. As for Fig~\ref{outbursts_all}, time $t=0$ is set at the peak of the observed flux, which may not be the true maximum of the outburst if the peak of the outburst has not been observed; this is the case in half of the cases presented here. 


The observed profiles have in common a short rise and less than 5~hr interval between quiescence and maximum, while more variability is visible during the decay; the initial decline appears to be rather steep with typically a drop by one magnitude in less than two hours, and no much variability appears from one outburst to another; the subsequent evolution shows more diversity, with, in some cases, tails lasting from five to ten hours and possibly secondary maxima.

These repetitive outbursts appear to be rare. AAVSO observations of V1223 Sgr have been available since 1992, and no outburst brighter than $V=12$ is present in the database with the exceptions of the 2014 outburst and of the 2020 series. The observations, however, have been sparse until the 2014 outburst and the 2018 low state. One should also note that V1223 Sgr has been most of the time at $V \simeq 13$, and that the series of outtburst occurred at the end of the recovery from a low state, when the source was still fainter than usual by about 0.5 -- 1 mag; no outburst was recorded after JD 2459120 when $V$ was brighter than 13.5.

\begin{table}
\caption{Observed properties of short outbursts of IPs detected in the AAVSO database} 
\label{tab}
\centering 
\begin{tabular}{l l l l l l l } 
\hline\hline 
Name &  JD & duration & $\Delta V$ & Comment\\
     &     &  (days)  &            &    \\
\hline 
FO Aqr & 2452540 & $<0.8$     & 1.8 & unspecified band\\
TV Col & 2455887 & 0.3 -- 0.9 & 1.6 \\
       & 2455910 & 0.5        & 1.2 \\
       & 2455913 & $<1.5$     & 1.2 \\
NY Lup & 2456847 & $>0.3$     & 1.3 & $\Delta V$ uncertain\\
       & 2457910 & 0.5        & 1.9 \\ 
EI UMa & 2457085 & 0.3        & 1.8 \\

\hline
\end{tabular}
\end{table}

   \begin{figure*}
   \centering
   \includegraphics[width=\textwidth]{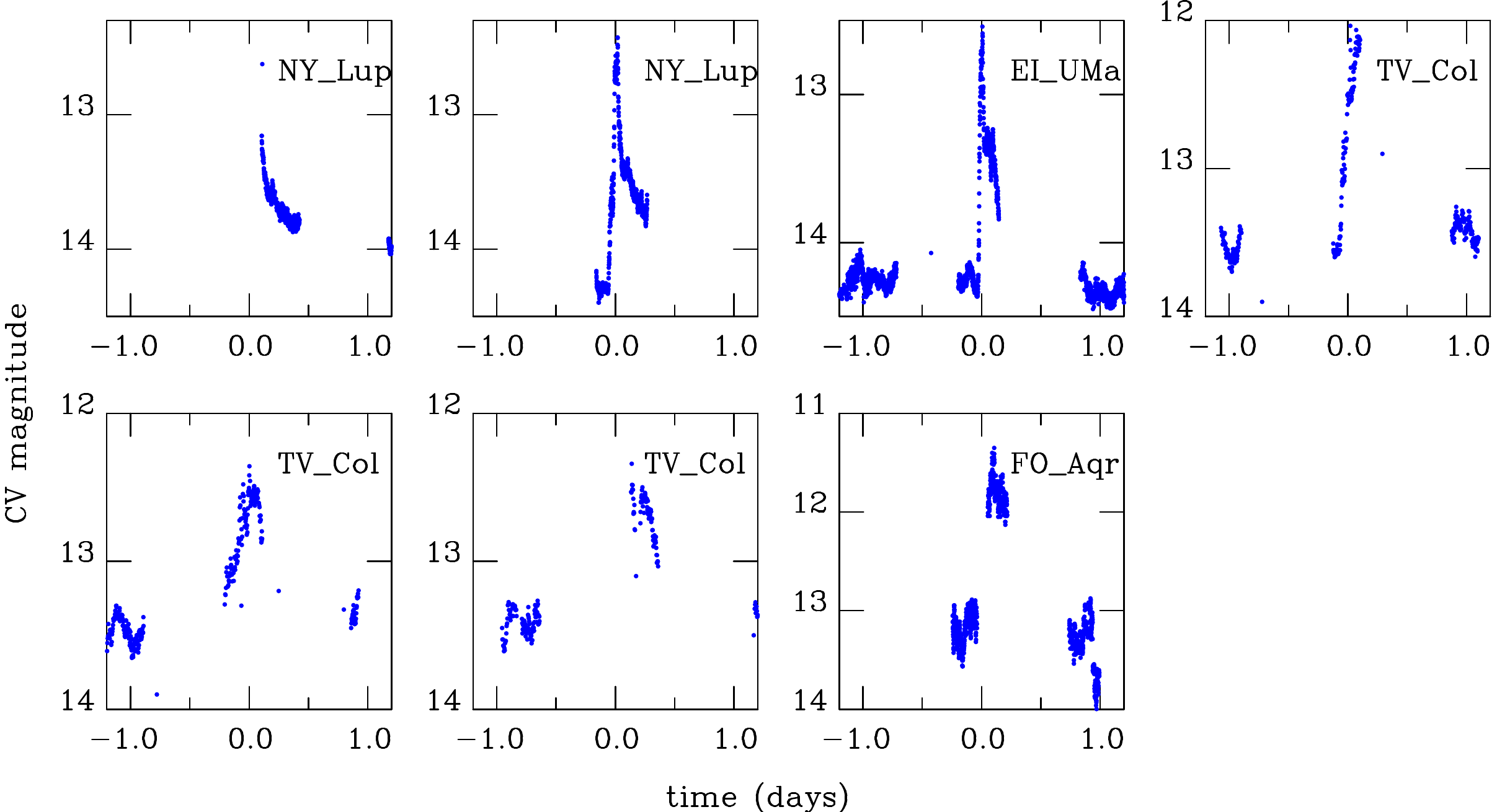}
   \caption{Short IP outbursts present in the AAVSO database.}
   \label{outbursts_IP}%
   \end{figure*}

\section{Short outbursts in other IPs}
\label{sec:short}

We explored the database of the AAVSO to search for short outbursts in known IPs. In the on-line version of the \citet{rk03} catalogue we selected the confirmed IPs  that are brighter than $V=16$ and searched for short (duration less than a day) outbursts with a magnitude decrease of at least $\Delta V = 1.5$ and for which a reasonable time profile was available. Apart from V1025 Cen, we did not find any other case of clusters of short outbursts, with the possible exception of TV Col that we discuss below.

We did find, however, isolated short outbursts in a few systems. Their characteristics are given in Table~\ref{tab} and the outburst time profiles are given in Fig. \ref{outbursts_IP}. In four cases, the outburst peak has been observed, so that the outburst amplitude, and, to a lesser extend, duration, is well determined. In the case of the first outburst of NY Lup, the maximum is set by a single observational point, much brighter than the neighbouring data points, which casts some doubts on the outburst amplitude. FO Aqr observations were performed using a CCD, but the band is not available.

TV Col was previously known as an IP that experienced several short outbursts; the other two sources for which short outbursts have been reported, CXOGBS J174954.5-294335 and XY Ari, are too faint to be routinely observed by amateur astronomers. To our knowledge, outbursts have not been reported in FO Aqr, NY Lup and EI UMa; a possible outburst from NY Lup  has been reported (vsnet-alert 24069\footnote{http://ooruri.kusastro.kyoto-u.ac.jp/mailarchive/vsnet-alert/24069}), but was not confirmed. The AAVSO database contains a significant number of observation sequences during which TV Col reached a magnitude brighter than 12.5, which probably correspond to outbursts, but these are usually visual observations, with large error bars, and the time coverage is sparse. Interestingly, TV Col had two outbursts that are separated by three days, occurring 33 days after the first outburst listed in Table~\ref{tab}. TV Col was observed during 5 to 7~h every night between the first two outbursts; it was also monitored after the third outbursts, but there are a few gaps in the data, up to five days; two upper limits (visual magnitude fainter than 12.8) exist in these gaps. Outbursts might thus have been missed, but it appears unlikely that a series of ten or more outbursts, as in the case of V1223 Sgr or V1025 Cen could have been missed. Nevertheless, a small series of outbursts might have been present. 

As in V1223 Sgr, the rises to outburst appear to be shorter than the decays, and the outburst durations are, when well constrained, significantly shorter than one day. Some outbursts, however, appear to have a very sharp rise as in the second outburst of NY Lup or that of EI UMa in Fig. \ref{outbursts_IP}.

Finally, we also found in the AAVSO data what appears to be a normal dwarf nova outburst that occurred in V1062 Tau; it lasted for about 4 days and the peak magnitude was recorded at JD 2459277; the increase in $V$ was 1.2. This source was previously not classified as a DN even though short and faint outbursts were reported \citep{lbo04}. We also found the end of a superoutburst that occurred in CTCV J2056-3014 at JD 2459100, which was followed by a reflare. The total duration of the outburst was at least 20 days. This source was already known as a DN, but this is the first detection of a superoutburst, which classifies this source as a SU UMa; this classification is confirmed by the identification of superhumps (vsnet-alert 24665\footnote{http://ooruri.kusastro.kyoto-u.ac.jp/mailarchive/vsnet-alert/24665}). A normal outburst was also found at JD 245983 which lasted for about 5 days. The outburst of V1026 Tau and the superoutburst of CTCV J2056-3014 are shown in the Appendix.

\section{Model} \label{model}

In order to interpret the short IP outbursts as magnetic--gating events we have first to adapt the \citet{ds10} model to the case of magnetized white dwarfs.
In this model, the instability arises when the inner disc radius is close to the corotation radius $r_{\rm c}$, defined as the radius at which the centrifugal forces on matter corotating with the white dwarf balance gravity forces:
\begin{equation}
r_{\rm c} = \left ( \frac{G M}{\Omega^2} \right)^{1/3},
\end{equation}
where $\Omega$ is the rotation frequency of the white dwarf. 

DS10 argue that the accretion disc interacts with the white dwarf magnetic field over a distance $\Delta r$ which is small (contrary to early assumptions of \citet{glp77,gl79a,gl79b} that \citet{w87} has later shown to be inconsistent).

DS10 define a characteristic accretion rate:
\begin{equation}
\dot{M}_{\rm c} = \frac{\eta \mu^2}{4 \Omega r_{\rm c}^5},
\end{equation}
where $\eta$ is a numerical factor describing the distortion of the magnetic field by the disc, which they take equal to 0.1, and $\mu$ is the magnetic moment of the white dwarf. $\dot{M}_{\rm c}$ corresponds to the rate at which the inner disc radius, defined at the magnetospheric radius, is equal to the corotation radius. If the inner disc radius is less than $r_{\rm c}$, accretion can proceed, and the standard disc equations should apply. In the opposite case, the accretion rate is vanishingly small. When centrifugal forces prevent accretion, DS10 assume that the surface density is such that the viscous angular momentum transport outside the interaction region is balanced by the angular momentum flux added by the magnetic field across the interaction region. They find:
\begin{equation}
\nu \Sigma = \frac{2}{3 \pi} \frac{\Delta r}{r_c} \left( \frac{r_{\rm c}}{r_{\rm in}} \right)^{9/2}  \dot{M}_{\rm c},
\label{bc0}
\end{equation}
where $\Delta r$ is the width of the interaction region.
Because the interaction region between the disc and the magnetosphere is small but finite, the transition between the two situations is smooth; DS10 use an interpolation functions of the form $(1 + \tanh x)$. The inner boundary condition $\nu \Sigma = 0$ at $r=r_{\rm in}$ where $\nu$ is the kinematic viscosity coefficient, $\Sigma$ the surface density and $r_{\rm in}$ the inner disc radius, then becomes:
\begin{equation} 
\label{bc1}
\nu \Sigma = \frac{2}{3 \pi} \frac{\Delta r}{r_c} \left( \frac{r_{\rm c}}{r_{\rm in}} \right)^{9/2}  \dot{M}_{\rm c} \left[ 1+\tanh \left( \frac{r_{\rm in}-r_{\rm c}}{\Delta r} \right) \right],
\end{equation}
which reduces to Eq. (\ref{bc0}) if the corotation radius is small and to $\nu \Sigma = 0$ when it is large. A similar approach for the position of the inner radius leads to:
\begin{equation}
\dot M  = -2 \pi r_{\rm in} \dot{r}_{\rm in} \Sigma + \frac{1}{2} \dot M_{\rm c} \left( \frac{r_{\rm c}}{r_{\rm in}} \right)^5 \left[ 1 - \tanh \left( \frac{r_{\rm in}-r_{\rm c}}{\Delta r_2} \right) \right],
\label{bc2}
\end{equation}
where $\dot{M}$ is the local mass transfer rate at the inner edge, calculated using Eulerian derivatives, which differs from the actual accretion onto the white dwarf by the term containing the time derivative of the inner radius $\dot{r}_{\rm in}$. $\Delta r_2$ is another characteristic length, which need not be equal to $\Delta r$, but is also assumed to be small as compared to $r_{\rm in}$. Equation (\ref{bc2}) reduces to $\dot M  = -2 \pi r_{\rm in} \dot{r}_{\rm in} \Sigma$ if $r_{\rm in} \gg r_{\rm c}$, i.e. accretion is impossible, and to $\dot M  = \dot M_{\rm c} (r_{\rm c}/r_{\rm in})^5$ if $r_{\rm in} \ll r_{\rm c}$. In the latter case, DS10 note that this sets the inner disc radius at a distance that differs, but not by a large amount, from the standard magnetospheric radius obtained by equating the magnetic pressure to the ram pressure of a spherically symmetric infalling matter.

$\Delta r$ and $\Delta r_2$ are thus the two parameters that describe the interaction of the magnetic field with the accretion disc; they are assumed to be small, but DS10 find that, in order to be stable with respect to the interchange stability, the interaction region cannot be too small; in the following, unless otherwise noted, we adopt their favourite values, namely $\Delta r/r_{\rm in} = 0.05$ and $\Delta r_2/r_{\rm in} = 0.014$.

The set of partial differential equations describing the time evolution of the accretion disc remains unaffected.

DS10 furthermore assume that the viscosity in the accretion disc varies as $\nu = \alpha (GM)^{1/2} (H/r)^2 r^{1/2}$, where $\alpha$ is the Shakura-Sunyaev parameter and $H$ is the disc scale height. They take $\alpha = 0.1$ and a constant $H/r=0.1$. They also assume that $\Delta r /r_{\rm in}$ and $\Delta r_2/r_{\rm in}$ take constant values. With these hypotheses, only three parameters determine the time evolution of the disc: the ratio $\dot{M}_{\rm tr}/\dot{M}_{\rm c}$, $\dot{M}_{\rm tr}$ being the mass transfer rate from the secondary, $\Delta r /r_{\rm in}$ and $\Delta r_2 /r_{\rm in}$. One should, in principle, add the outer disc radius $r_{\rm out}$, but, as we shall see later, the solution is almost independent on $r_{\rm out}$ provided that $r_{\rm out}$ is not too close to $r_{\rm in}$. The results from DS10, DS11 and DS12, although obtained in the context of low-mass X-ray binaries and young stars, should therefore be applicable to IPs.

Some of the above mentioned assumptions are unnecessary, and questionable in the case of IPs. In particular, the power-law dependence of the viscosity and the assumption $H/r=0.1$ are not well justified in cataclysmic variables\footnote{Not that they are more justified in accretion discs in other systems.}. We therefore used our code for solving the time-evolution equations of accretion discs \citep{hmd98}, with  modified boundary conditions.

We first checked that we do obtain the same results as DS10 when making the same assumptions, which is indeed the case, with the notable difference that the inner radius and mass accretion rate vary five times more rapidly than in DS10, although the shape and amplitude of these variations are identical. We also find the same instability zones, and in particular the two instability regions labelled RI and RII by DS12, RI being located at low mass transfer rates for a relatively narrow range of $\Delta r_2/r_{\rm in}$ and RII being located at mass transfer rates close to $\dot M_{\rm c}$ for a broad range of $\Delta r_2/r_{\rm in}$. We checked that the disc stability as well as the outcome of the instability do not depend on the outer disc radius when it is large; there was no difference between $r_{\rm out} = 400 r_{\rm in}$ and $r_{\rm out} = 40 r_{\rm in}$. These large $r_{\rm out} / r_{\rm in}$ are appropriate for young stars and low-mass X-ray binaries; this is not the case for IPs in which this ratio is a few, and when we considered $r_{\rm out} = 3 r_{\rm in}$, significant deviations from the large $r_{\rm out} / r_{\rm in}$ case were found, such as for example a shortening of the instability period by a factor 1.5.

We also noted that the outcome of the instability depends somewhat on the details of the assumed interaction between the magnetosphere and the accretion flow that enter in the boundary condition. For example, replacing $\Delta r / r_{\rm in}$ by $\Delta r / r_{\rm c}$ in Eq. \ref{bc1} results in stability zones that are significantly different, even though $r_{\rm c}$ and $r_{\rm in}$ do not differ by more than a few percent.

We then solved the full disc equations for a system with parameters appropriate for V1223 Sgr: orbital period $P_{\rm orb}= 3.36$~h, spin period $P_{\rm spin} = 0.207$~h, primary mass $M_1= 0.9$~M$_\odot$, secondary mass $M_2=0.33$~M$_\odot$. 

  \begin{figure*}
   \centering
   \includegraphics[width=0.99\columnwidth]{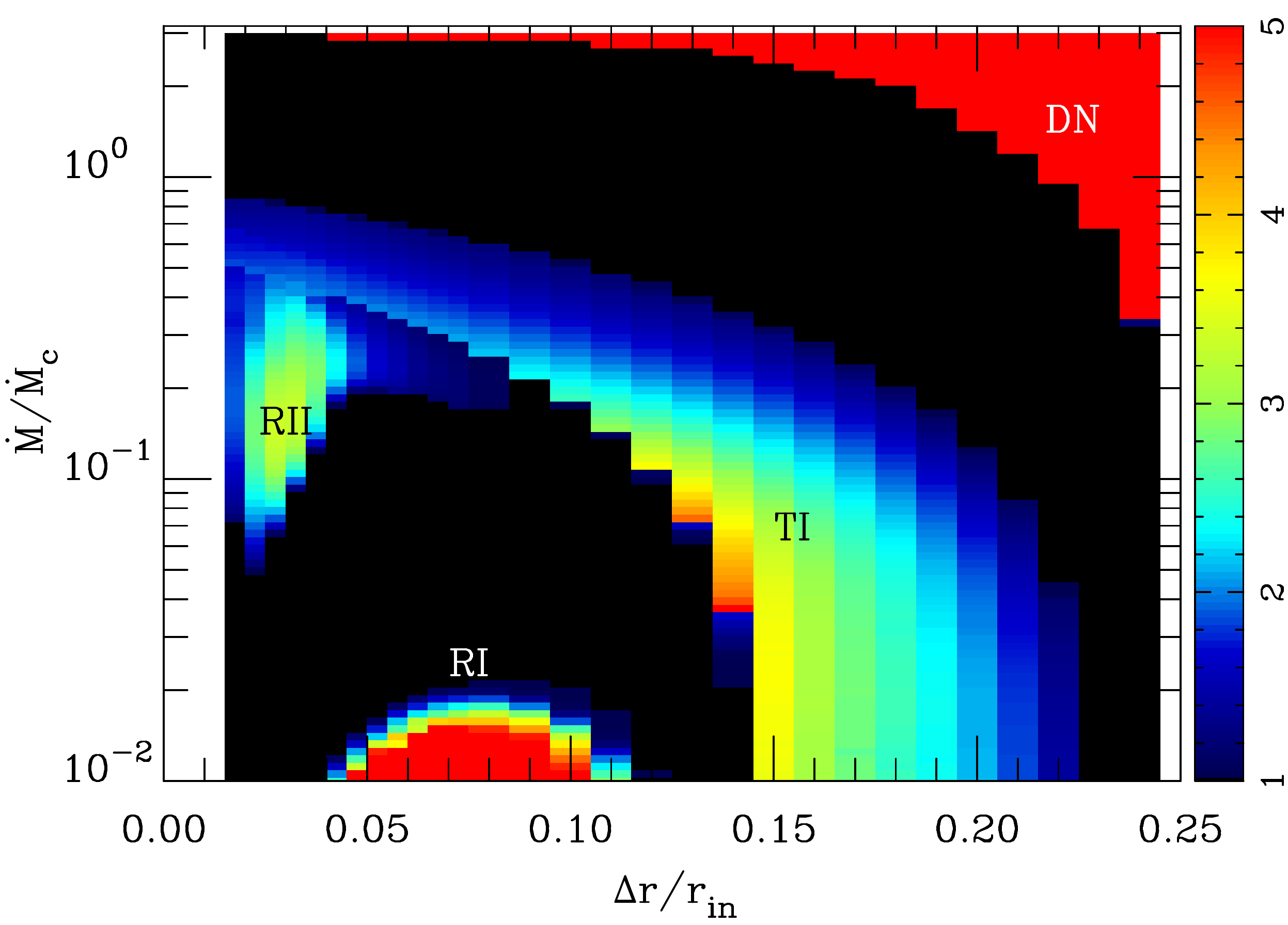}
   \hfill
   \includegraphics[width=0.99\columnwidth]{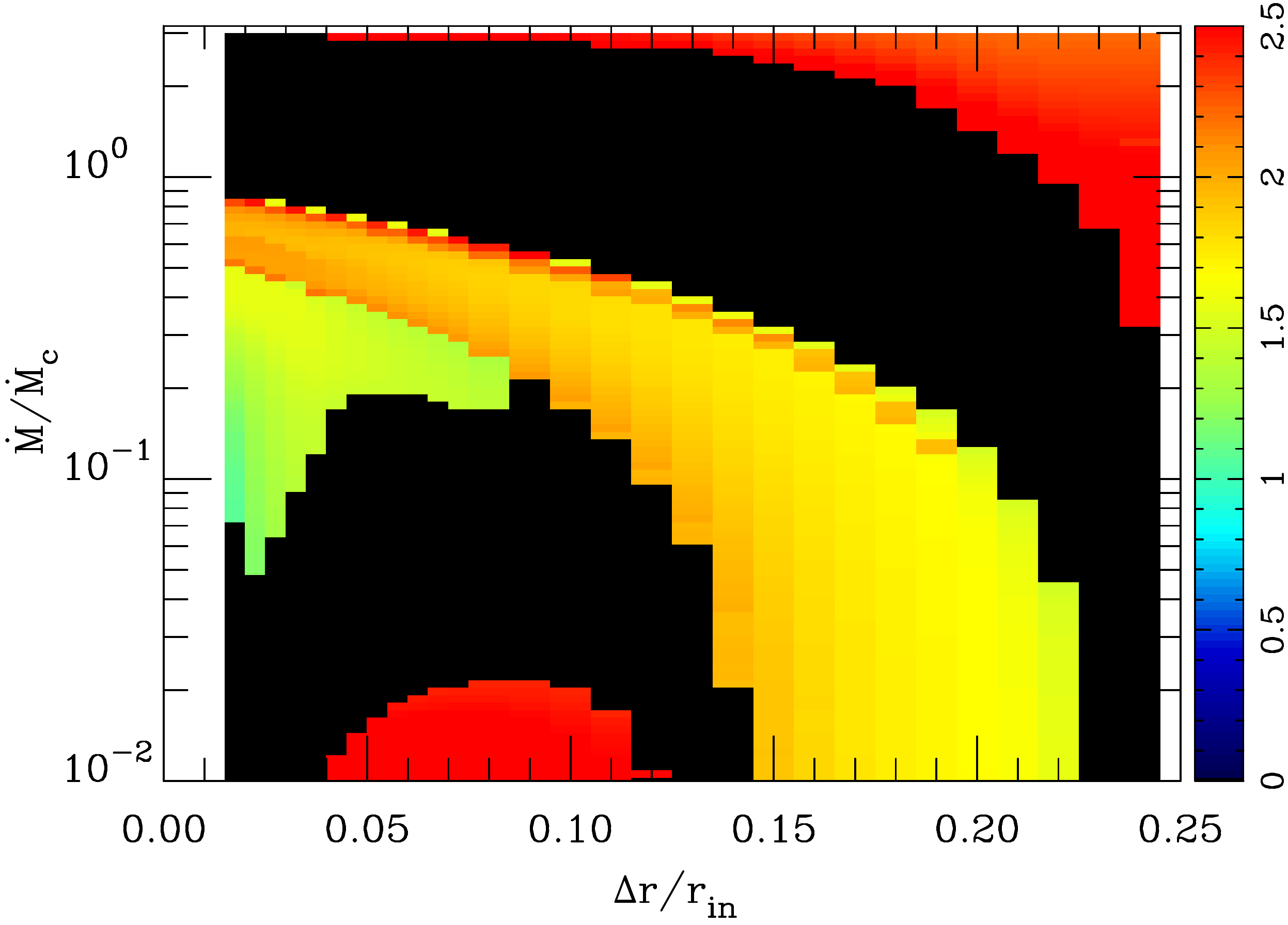}
   \caption{Instability map in the cold disc case for $\mu_{33}=1.5$. {\it Left:} outburst amplitude defined as $A = \dot M_{\rm max}/\dot M_{\rm tr}$. Here, $\dot M_{\rm c} = 2.8 \times 10^{16}$~g~s$^{-1}$. {\it Right:} logarithm of the recurrence time measured in days. Black areas correspond to stability. Regions labelled 'RI' and 'RII' correspond to the magnetic gating instability; the dwarf nova instability occurs in the region labelled 'DN', and a thermal instability due to the dissociation of the hydrogen molecule occurs in region labelled 'TI'.}
   \label{stab_cold}%
   \end{figure*}

Since systems exhibiting short outbursts do not have normal dwarf nova outbursts, they must be stable with respect to the thermal-viscous instability. This implies that either the mass transfer is low, with $\dot M_{\rm tr}$ less than the maximum rate for being on the cold, stable branch at the inner edge of the disc, $\dot M_{\rm tr} < \dot M_{\rm crit}^-(r_{\rm in})$, or that it is larger than the minimum rate for being on the hot branch estimated at the outer radius, $\dot M_{\rm tr} > \dot M_{\rm crit}^+(r_{\rm out})$. This requirement does not mean that individual systems cannot oscillate between high and low states without experiencing normal dwarf nova outbursts. This is possible if for a disc initially sitting on the hot branch, the disc disappears before the mass transfer rate gets below $\dot M_{\rm crit}^+(r_{\rm out})$ as mentioned in \citet{hl17b} and as was apparently observed in the case of FO~Aqr \citep{lgk20}.

The magnetic gating instability can exist in principle for very low mass accretion rates, with $\dot{M}_{\rm tr}/\dot{M}_{\rm c}$ as low as $10^{-5}$; the requirement that the disc is cold everywhere is therefore not incompatible with the criterion for the magnetic gating to be effective; this is, however, not true for hot discs, because the magnetic gating instability does not exist for $\dot{M}_{\rm tr}/\dot{M}_{\rm c}$ larger than about unity. We examine these two cases in turn, after discussing conditions under which a disc forms.

\subsection{Existence of a disc}

The condition for the formation of a disc is, in the general case, $r_{\rm circ} > \max (r_*, r_{\rm mag})$ \citep{fkr02}, where $r_{\rm circ}$ is the circularization radius, i.e. the radius at which matter leaving the secondary star would form a circular orbit around the primary, $r_*$ is the primary's radius, and $r_{\rm mag}$ the magnetospheric radius. This condition, however, does not apply when $r_{\rm circ}$ is of order of or slightly larger than $r_{\rm c}$. When this happens, matter cannot enter the magnetosphere because of centrifugal forces, and cannot be ejected from the system either because the velocity acquired by matter spun up at the magnetosphere is less that the escape velocity, unless $r_{\rm mag}  > 1.26 r_{\rm c}$ (DS10). Therefore, if $r_{\rm c} < r_{\rm mag} < 1.26 r_{\rm c}$, a disc forms that cannot initially accrete onto the white dwarf. Matter piles up, the inner disc radius decreases until it becomes equal to $r_{\rm c}$ (to within $\Delta r$), at which point accretion begins. A steady configuration is possible because there is a transfer of angular momentum from the primary to the accretion disc at its inner edge; this is not possible when the white dwarf rotates slowly and $r_{\rm c}$ is large, in which case the condition $r_{\rm circ} > \max (r_*, r_{\rm mag})$ still applies.

\subsection{Cold discs}

We assume here a magnetic moment $\mu = 1.5 \times 10^{33}$~G~cm$^3$, hence $r_{\rm c}=1.2 \times 10^{10}$cm and $\dot M_{\rm c}=2.1 \times 10^{16}$~g~s$^{-1}$. Our magnetic moment differs from the estimate by \cite{nws04} which does not apply because the white dwarf spin is not in equilibrium. When taking $\Delta r/r_{\rm in} = 0.05$ and $\Delta r_2/r_{\rm in} = 0.014$ as in DS10, we find that the disc is unstable if the mass transfer rate $\dot M_{\rm tr}$ is less than $0.015 \dot M_{\rm c} = 4.1 \times 10^{14}$~g~s$^{-1}$, which corresponds to instability zone RI, and for the range $6 \times 10^{15} < \dot M_{\rm tr} < 1.0 \times 10^{16}$~g~s$^{-1}$, or $0.2 < \dot M_{\rm tr}/\dot M_{\rm c} < 0.36$. These values differ somewhat from the values that we obtain when using the simplified disc model used by DS10, which are $\dot M_{\rm tr}/\dot M_{\rm c} < 0.124$ and $0.97 < \dot M_{\rm tr}/\dot M_{\rm c} < 1.20$. There are two main reasons for this difference. First, the viscosity does not scale as $r^{1/2}$, but rather slightly decreases with radius; second, the outer disc radius $r_{\rm out}$ is not large as compared to the inner radius: $r_{\rm out}/r_{\rm in}$ is only 2.35. For these reasons, the periodicity of the instability is also shortened as compared to the simplified case; for $\dot M_{\rm tr}/\dot M_{\rm c} = 0.072$, the time interval between outbursts is reduced by a factor 2.5.

Figure \ref{stab_cold} shows the outburst amplitude in a $(\Delta r/r_{\rm in},\dot{M_{\rm tr}}/\dot M_{\rm c})$ plane. Four instability regions appear. The strong instabilities that appear at large $\dot M_{\rm tr}/\dot M_{\rm c}$, of order unity or larger, correspond to the normal dwarf nova instability. The instability threshold is larger than $\dot M_{\rm crit}^-(r_{\rm in})$, because of the stabilizing effect of the boundary condition. The boundary condition expressed by Eq. (\ref{bc2}) damps rapid variations in the mass accretion rate because of its strong coupling with $r_{\rm in}$, which more than compensates the increase in $\Sigma$ with respect to the standard $\nu \Sigma =0$ condition. It is also worth noting that the unstable domain depends on $\Delta r$ for the same reason: the smaller $\Delta r$, the stronger the coupling between inner-edge $\dot M$  and $r_{\rm in}$. A second, much weaker unstable region is found at lower mass transfer rates and covers almost the full range of $\Delta r$. This region corresponds to a thermal instability that develops when the S-curves describing the thermal structure of the accretion disc show little wiggles due, in general, to the dissociation of hydrogen molecules. Heating and cooling fronts propagate back and forth but do not cause strong instabilities because the viscosity parameter remains constant. These two zones are approximately parallel; the dependence on $\Delta r$ should not come as a surprise because the inner boundary condition affects the whole disc as $r_{\rm out} /r_{\rm in}$ is only a few. The other two regions correspond to the magnetic gating instability. The region centered on $\Delta r/r_{\rm in}$ = 0.03 and $\dot{M_{\rm tr}}/\dot M_{\rm c} = 0.2$ is the second instability region (RII) in DS12; the region at $\dot{M_{\rm tr}}/\dot M_{\rm c} < 0.02$ and $\Delta r/r_{\rm in} \sim 0.07$ is labelled RI in DS12.

The time-variations of the accretion rate can be complex. Whereas low amplitude variations (typically by a factor two or less) are almost sinusoidal, complex shapes consisting of multi-peaks periodic profiles are found when the instability is strong, and in particular when the magnetic gating instability couples with the thermal instability. Here, we define the recurrence time as the periodicity of the $\dot M$ variations, implying that the time interval between two outbursts may be shorter than the recurrence time. Transitions between simple and multiply peaked light curves correspond to a discontinuity of the recurrence time in Fig. \ref{stab_cold}. One must also note that when the amplitude of the magnetic gating instability is strong, recurrence times are large.

We did not explore the effect of varying $\Delta r_2$ on the instability outcome. Judging from the DS12 results, varying $\Delta r_2$ does have an influence on the instability amplitude and period, but it is significantly less than that of $\Delta r$ variations (see their figures 6 and 7). DS12 also find that changing $\Delta r_2$ does not alter the conclusion that short periods are obtained only when the amplitudes are low.

  \begin{figure*}
   \centering
   \includegraphics[width=0.99\columnwidth]{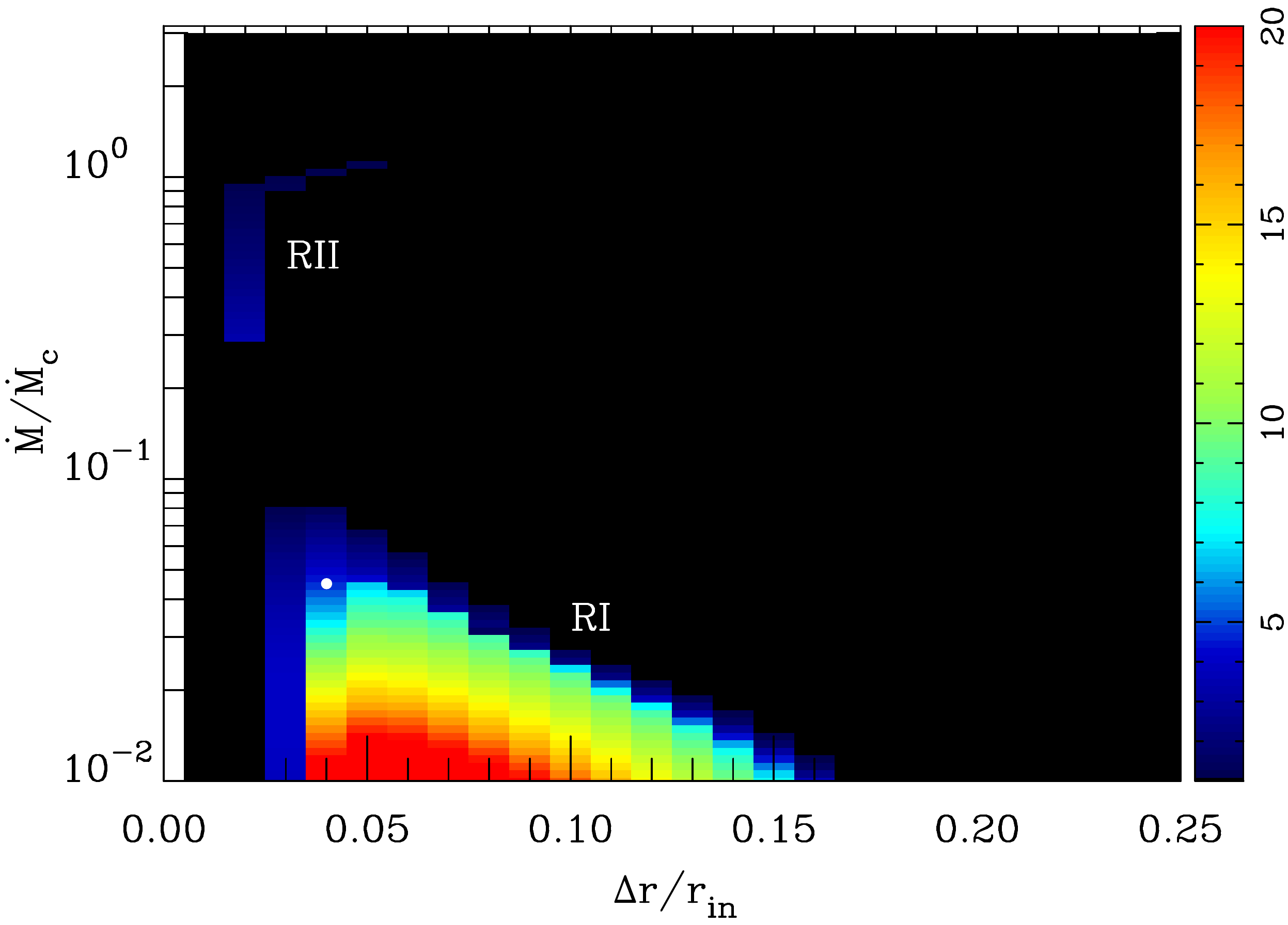}
   \hfill
   \includegraphics[width=0.99\columnwidth]{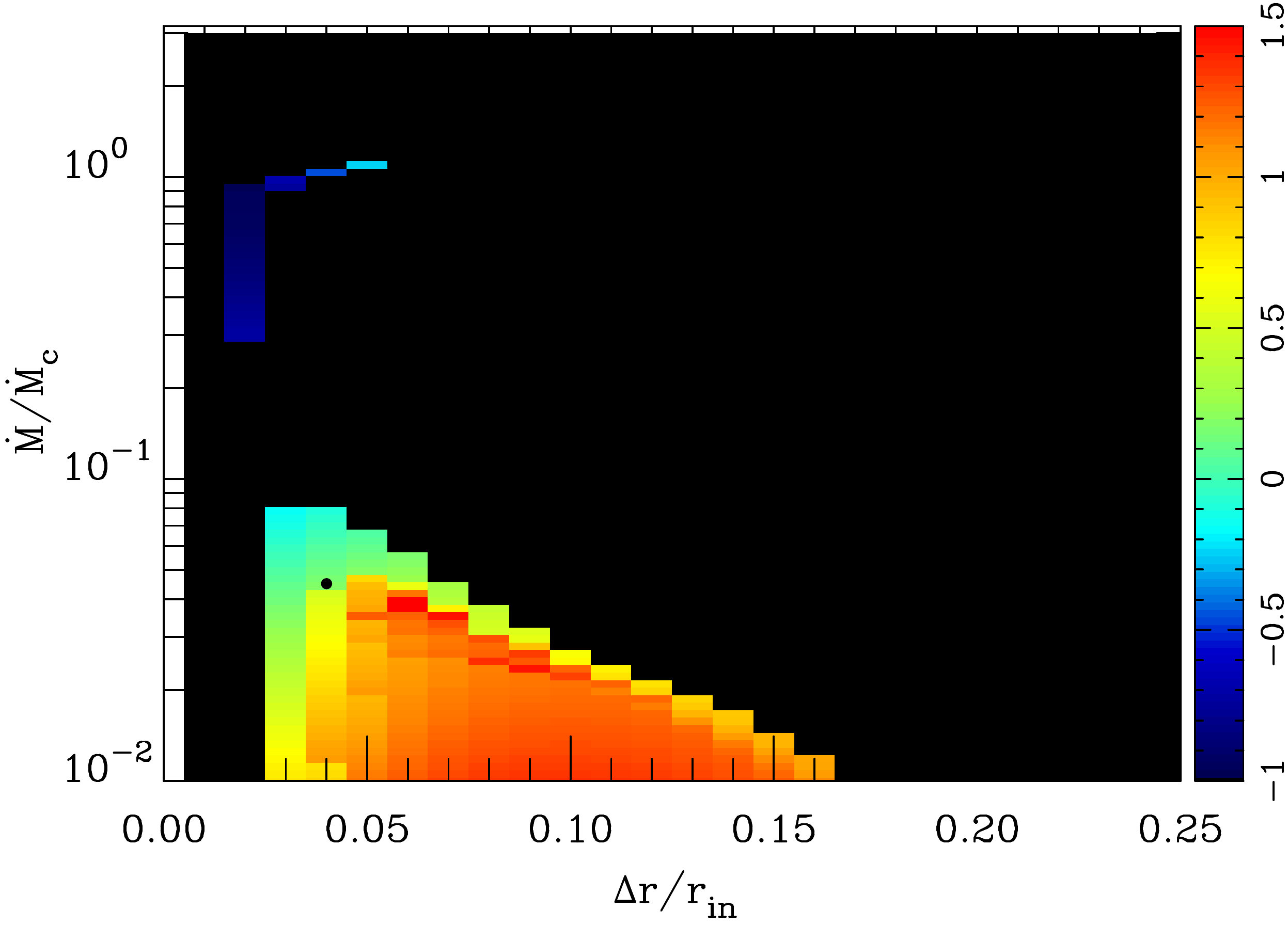}
   \caption{Instability map in the hot disc case for $\mu_{33}=15$. {\it Left:} outburst amplitude defined as $A = \dot M_{\rm max}/\dot M_{\rm tr}$; here, $\dot M_{\rm c} = 2.8 \times 10^{18}$~g~s$^{-1}$. {\it Right:} logarithm of the recurrence time measured in days. Regions labelled 'RI' and 'RII' correspond to the magnetic gating instability with the same label as in DS10. The white (left panel) and black (right panel) dots indicate the position of the light curve shown in Fig. \ref{lcurve}}
   \label{stab_hot}%
   \end{figure*}

The magnetic gating can therefore develop in intermediate polars with cold discs; however, this instability is not very strong, with amplitudes less than typically a factor two except for very low mass transfer rates, of order of a few times 10$^{14}$~g~s$^{-1}$ or less (RI), but the recurrence time is then very long (a year or more). Shorter recurrence times, of order of days to weeks, are found in the RII regime, but the amplitude is then low, implying that the fluctuations of the accretion rate are close to being sinusoidal;  the outburst duration is then longer than a few days. In summary, short outbursts cannot be produced by the magnetic gating instability in cold discs.

   \begin{figure}
   \includegraphics[width=\columnwidth]{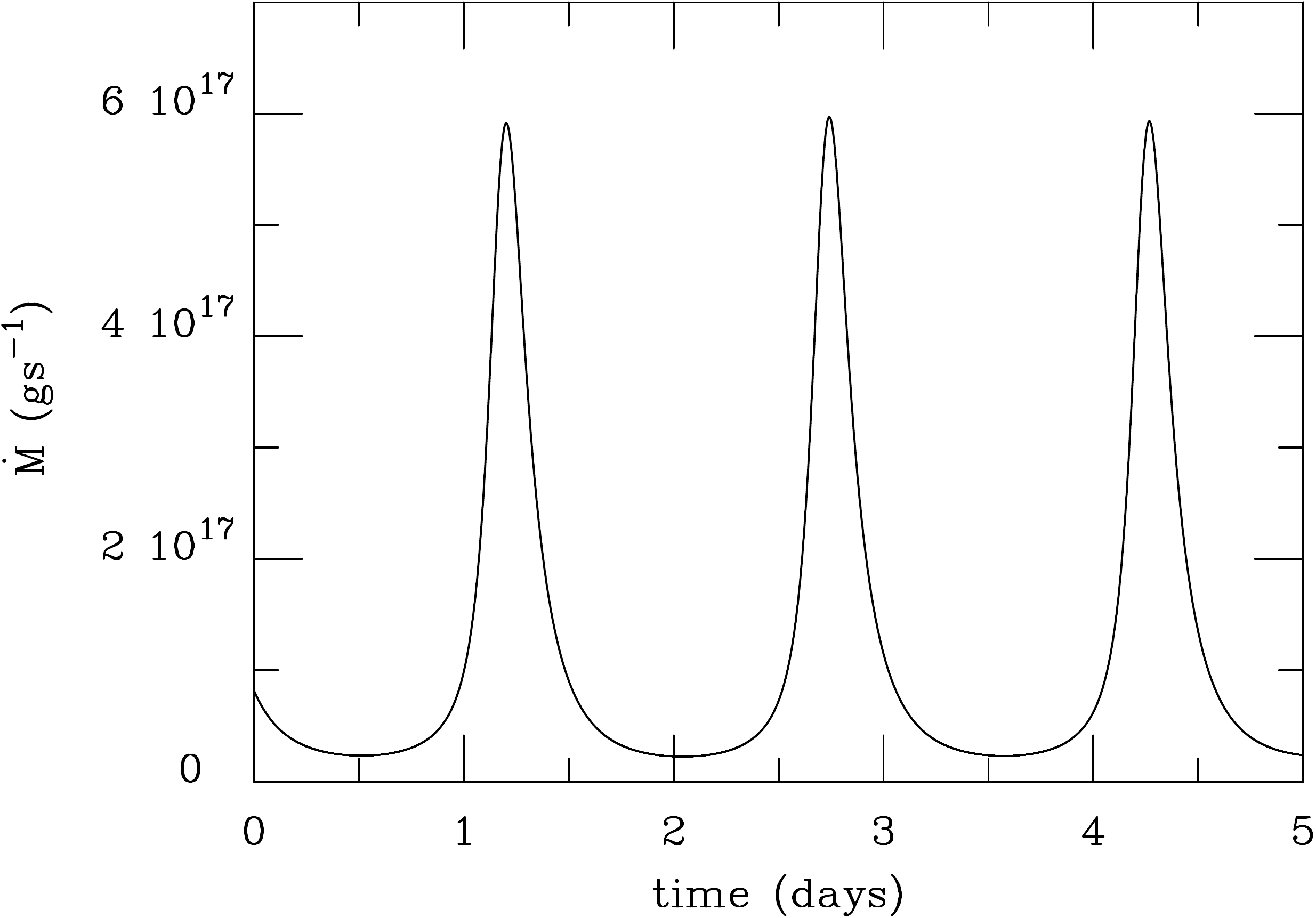}
   \caption{Example of light curve obtained with $\mu_{33}=15$, $\Delta r/r_{\rm in} = 0.04$, $\Delta r_2/r_{\rm in} = 0.014$ and $\dot M_{\rm tr} = 1.24 \times 10^{17}$~g~s$^{-1}$.}
   \label{lcurve}
   \end{figure}
   
   \begin{figure}
   \includegraphics[width=\columnwidth]{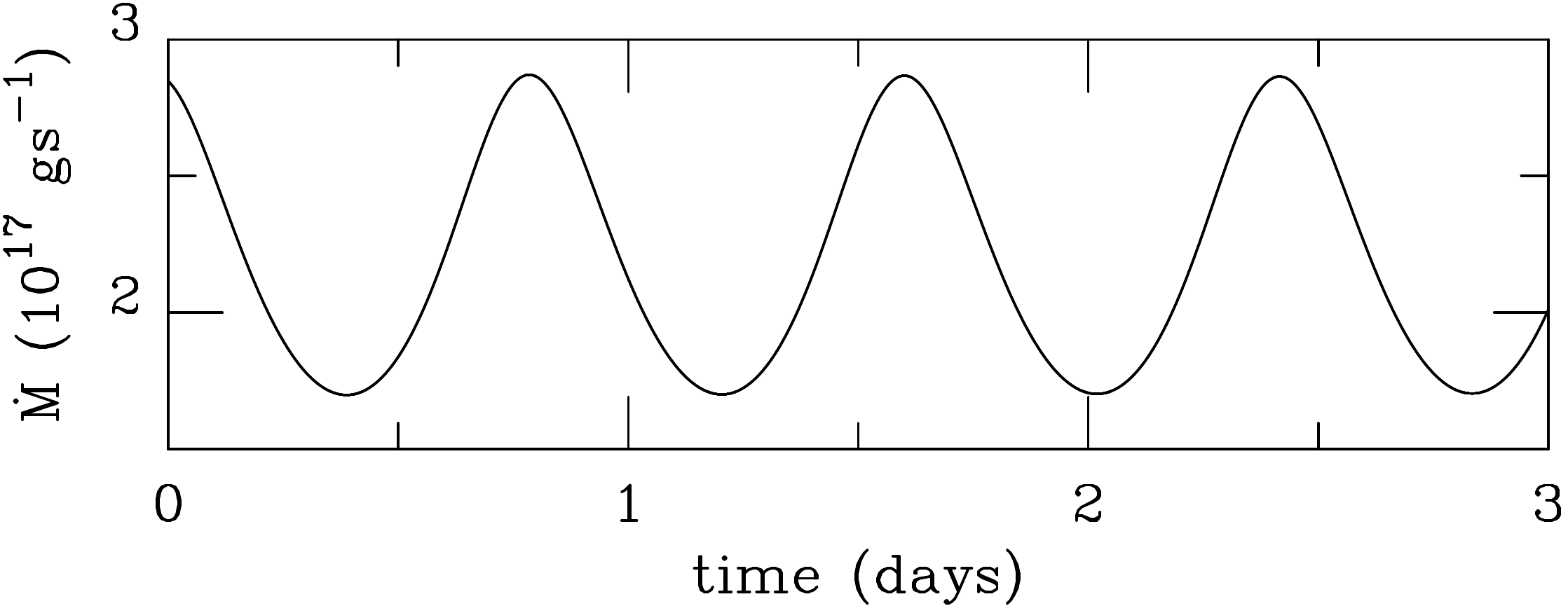}
   \includegraphics[width=\columnwidth]{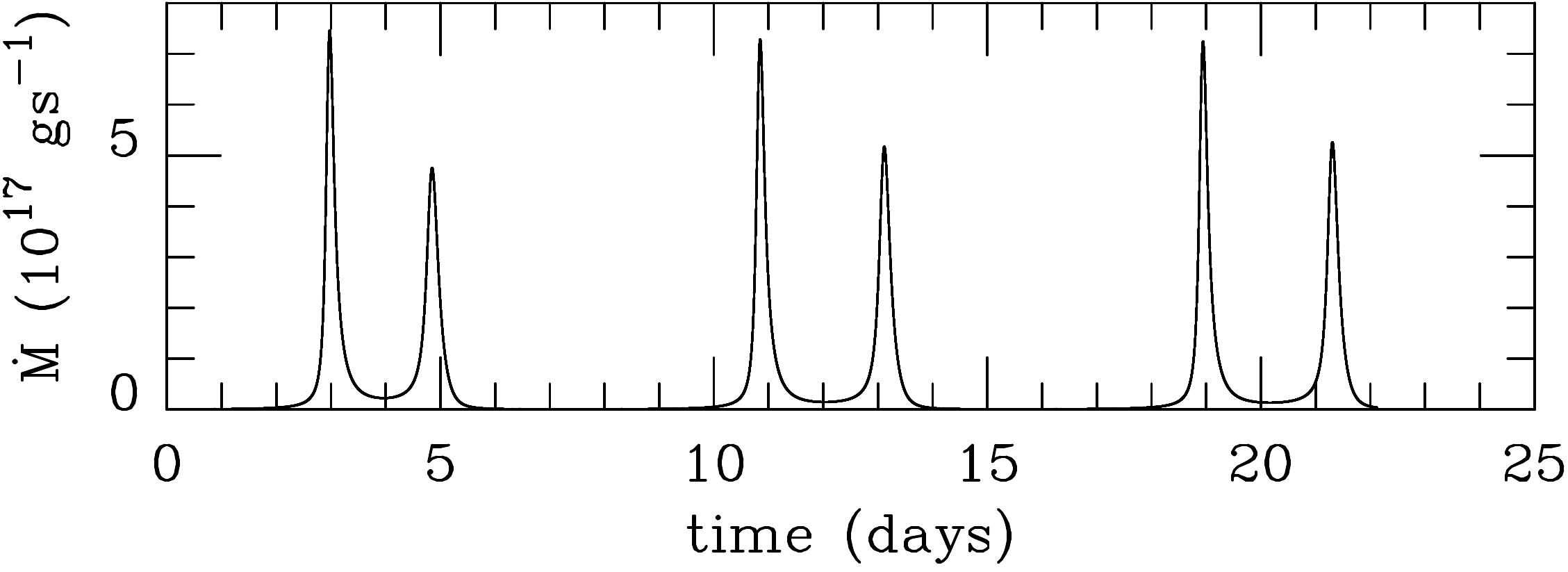}
   \includegraphics[width=\columnwidth]{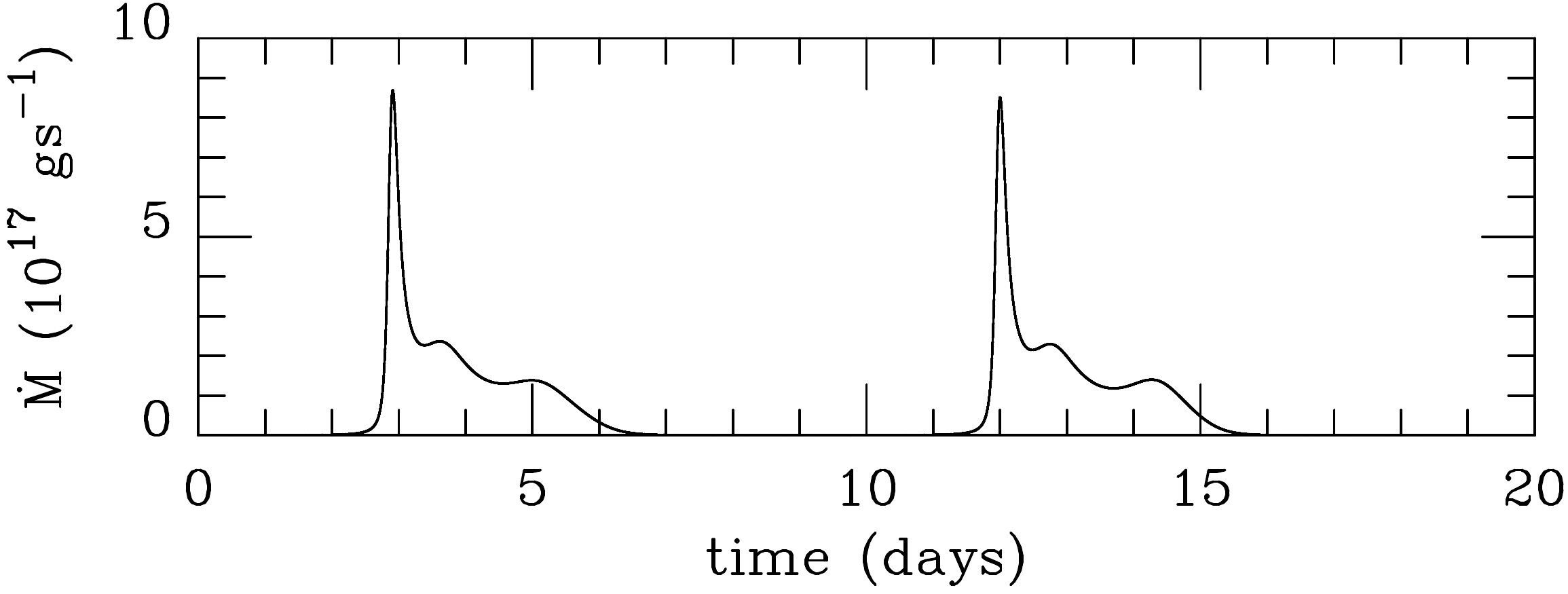}
   \caption{Additional examples of light curves obtained with $\mu_{33}=15$ and $\Delta r_2/r_{\rm in} = 0.014$: {\it top:} $\Delta r/r_{\rm in} = 0.04$ and $\dot M_{\rm tr} / \dot M_{\rm c} = 0.080$. {\it Intermediate:} $\Delta r/r_{\rm in} = 0.04$ and $\dot M_{\rm tr} / \dot M_{\rm c} = 0.016$. {\it Bottom:} $\Delta r/r_{\rm in} = 0.05$ and $\dot M_{\rm tr} / \dot M_{\rm c} = 0.025$.}
   \label{lc2}
   \end{figure}

\subsection{Hot discs}

The disc is hot and stable with respect to the thermal-viscous instability if the mass transfer rate is larger than $\dot M_{\rm crit}^+(r_{\rm out})$. On the other hand, $\dot M_{\rm tr}$ has to be less than about $\dot M_{\rm c}$ for the magnetic gating instability to be effective, which leads to condition for the latter to operate:
\begin{equation}
8.07 \times 10^{15} r_{\rm out,10}^{2.64} M_1^{-0.89} < \dot M_{\rm tr} < 2.63 \times 10^{14} P_{\rm sp,hr}^{-7/3} M_1^{-5/3} \mu_{33}^2,
\end{equation}
where we have used the analytic fits for $\dot M_{\rm crit}^+$ provided by \citet{ldk08}, neglecting the very weak dependence on $\alpha$. Here, $\dot{M}_{\rm tr}$ is measured in g~s$^{-1}$, $r_{\rm out,10}$ is the outer disc radius measured in $10^{10}$ cm, $P_{\rm sp,hr}$ is the white dwarf spin period measured in hours, and $\mu_{33}$ is the magnetic moment in units of $10^{33}$G. This condition is satisfied if the magnetic moment is larger than a critical value:
\begin{equation}
\mu_{33,\rm crit} = P_{\rm sp,hr}^{7/6} M_1^{0.39} r_{\rm out,10}^{1.32}
\label{mu_min}
\end{equation}
For the parameters of V1223 Sgr, this requires $\mu_{33} > 3.6$. In order to be able to explore a large domain in $\dot M_{\rm tr}$, we we assume in the following $\mu_{33} = 15$. Such a large $\mu_{33}$ would be more appropriate for polars, but is required if one wishes to consider mass transfer rates as low as $10^{-2}\dot M_{\rm c}$.

Figure \ref{stab_hot} shows the unstable domains for hot discs. As in the cold case, we find two well separated instability regions for the magnetic gating instability, one at low mass transfer rates, and one for high mass transfer rates. The unstable region does not extend much below $\dot{M}_{\rm tr}/\dot{M}_{\rm c} = 10^{-2}$ because the host disc then becomes subject to the thermal-viscous instability. In fact, for $\Delta r/r_{\rm in} = 0.1$, the disc is unstable at $\dot{M}_{\rm tr}/\dot{M}_{\rm c} = 10^{-2}$. The thermal-viscous instability is triggered at lower $\dot M_{\rm tr}$ than in the non-magnetic case, because the boundary condition (\ref{bc1}) results in a larger $\Sigma$ throughout the disc; the non-magnetic and magnetic steady-state solutions deviate by a term of order of $1-(r_{\rm in}/r)^{1/2}$ which is significant everywhere because $r_{\rm out}/r_{\rm in}$ is at most of order of a few. In the case shown in Figs \ref{stab_hot}, $\dot M_{\rm crit}^+/\dot M_{\rm c}$ is 0.05, meaning that the critical mass transfer rate for the thermal-viscous instability is reduced by a factor 5. 

As in the cold disc case, the unstable domain with high mass transfer rates has low amplitudes and short recurrence times; it corresponds to the region labelled RII in DS12. In this regime, the accretion rate varies smoothly by less than a factor two with a period of a few hours; this requires $\dot M_{\rm tr}$ larger than about $10^{18}$~g~s$^{-1}$ for $\mu_{33}=15$, and this limit scales as $\mu^2$. 

On the other hand, the regime at low mass transfer rates, with $\dot M_{\rm tr} < 5 \times 10^{16}$~g~s$^{-1}$ is unstable if $\Delta r/r_{\rm in}$ is in the range 0.05 - 0.10, resulting in large amplitudes, of the order of 10 -- 20, and relatively short periodicities. These correspond well to the sequence of short outbursts observed in V1223 Sgr. As an illustration, Fig. \ref{lcurve} shows the variations of the mass accretion rate when $\dot M_{\rm tr} = 0.045 \dot M_{\rm c} = 1.24 \times 10^{17}$~g~s$^{-1}$, $\Delta r/r_{\rm in} = 0.04$. These parameters correspond to a point in the instability map (Fig. \ref{stab_hot}) located at about half the maximum unstable mass transfer rate in the RI region. The light curve shows outbursts with a FWHM of 5.1~h and a recurrence rate of 1.5~d. Since our simulations show that the disc is still stable on the hot branch for mass transfer rates 5 times smaller than considered here, one would obtain similar results with $\mu_{33}=7$, which is more appropriate for an IP. For reference, Fig.\ref{lc2} shows a few additional examples of light curves that we obtained. In all cases, the outbursts FWHM is a few hours. As mentioned above, sinusoidal light curves are obtained only when the system is close to stability and only for a narrow range of parameters. Large amplitudes are obtained for low mass transfer rates, which require large magnetic moments for the disc to be on the hot state, and are thus unlikely to occur in real systems.

We can now revisit the minimum $\mu$ required for the magnetic gating instability to be effective in a hot disc. We assumed that the limiting $\dot M_{\rm tr}/M_{\rm c}$ was unity, while our simulations show that if is of order of 0.05; on the other hand, we overestimated by a factor 5 the minimum mass transfer for the disc to be hot and thermally stable. Correcting for these, we now get:
\begin{equation}
\mu_{33, \rm crit} = 8 P_{\rm sp,hr}^{7/6} M_1^{0.39} r_{\rm out,10}^{1.32}.
\label{mu_min1}
\end{equation}

One should note that the outer disc radius can be smaller than the tidal truncation radius if the disc mass is not large and if the mass transfer rate is not constant but steadily increases. This occurs in particular when the system emerges from a low state, and it might explain why sequences of short outbursts were observed in V1223 Sgr as well as in V1025 Cen precisely when the system was getting back to a normal state after a low state. This would allow lower magnetic moments and mass transfer rates.

\subsection{V1025 Cen}

The recurrence and duration of V1025 Cen outbursts \citep{llh22} are similar to those of V1223 Sgr; one therefore requires the same parameters as for V1223 Sgr; in particular $\dot M_{\rm tr}/\dot M_{\rm c} \simeq 0.05$. V1025~Cen has an orbital period of 85 min, and a spin period of 36 min \citep{bcr98,hbb98}. The corotation radius is thus $r_{\rm c} = 2.5 \times 10^{10} M_1^{1/3}$~cm. The primary mass is very uncertain; \citet{sdw19} find $M_1 = 0.61 \pm 0.14$~M$_\odot$ from X-ray observations; here, we take $M_1 = 0.8$~M$_\odot$, corresponding to the average primary mass in IPs. The outer disc radius can reach the spherical Roche radius of the primary which, for a mass ratio of 0.05, is $2.68 \times 10^{10}$~cm, only slightly larger than $r_{\rm c} = 2.32 \times 10^{10}$ ~cm. This makes the direct applicability of the model questionable, because $r_{\rm in} \approx r_{\rm c}$ is so close to $r_{\rm out}$ that the accretion flow forms rather a ring (torus) than a proper disc. \citet{hwb02} find that phase-resolved spectroscopy observations of V1025 Cen cannot be readily explained in either the classical disc-fed, or stream-fed, or even disc-overflow scenarios, making the nature of the accretion flow quite uncertain. The magnetic gating may still operate, but the predictions of the model in its present form cannot be taken at face value. Similarly, the stability of the disc with respect to the thermal-viscous instability, which sets the minimum $\dot M_{\rm tr}$ is also questionable. This said, the light curve in Fig. \ref{lcurve} looks very much like the one observed in V1025 Cen \citep[see Fig. 3, in][]{llh22}.

\section{Conclusions}
\label{sec:concl}

We have shown that the sequence of short outbursts discovered by \citet{csk22} in V1223 Sgr, lasting for a few hours and recurring every few days, is very similar to that observed in V1025 Cen by \citet{llh22}. Exploring the AAVSO database, we found that TV Col might have had several closely spaced short outbursts, but the data coverage is insufficient to ascertain this. 

These outbursts are similar in duration and amplitude to the isolated outbursts found in a few intermediate polars, and in particular TV Col. An examination of the AAVSO data base lead us to spot three new IPs that showed short isolated outbursts: FO Aqr, NY Lup and RI UMa. As by-product of this analysis, we found a normal DN outburst in V1062 Tau, which was not previously known as a DN, and a superouburst in CTCV J2056-3014, making it the third IP member of the SU UMa class. 

These outbursts cannot be due to the thermal-viscous instability of accretion discs which accounts for the dwarf nova phenomenon, because both the recurrence time and the outburst duration are too short. Instead we show that the series of short bursts observed in V1223 Sgr can be explained by the magnetic--gating instability model proposed by \citet{st93} and further developed by DS10 and DS12 in the context of LMXBs and young stars. We have adapted their model to the context of IPs. We found that the properties of the magnetic instability are qualitatively similar when using realistic models of accretion discs and when using simplified models with power-law variations, but both the unstable regions, the recurrence time and the amplitude of fluctuations may differ significantly. We argue that, judging from the light curve, the magnetic--gating instability should be also the cause of V1025 Cen series of outbursts but in this case the model should be adapted to an accretion torus, since the presence of an accretion disc in this system is unlikely.

We found that the magnetic gating instability may exist in cold discs, but the quiescence intervals and the outburst duration are far too long to account for the observations. When the entire disc is hot and stable with respect to the thermal-viscous instability, it can be subject to the magnetic gating instability if the magnetic field of the white dwarf is strong enough.

Whether isolated outbursts can be due to the magnetic gating instability remains an open question. The magnetic gating instability can produce long recurrence times, of the order of a month or possibly more; however, the outburst duration would then also be long, because the ratio between the outburst  recurrence time and duration is typically the ratio between the mean mass transfer rate and peak accretion rate, that does not exceed 30 for the cases considered here. 

It is also interesting to note that the time profiles of isolated outbursts sometimes have a sharp rise and a slower decay, in contrast with the almost symmetric profiles observed in V1223 Sgr and V1025 Cen.

We finally note that our models rely on a stability analysis of the steady state. This is, however, not a well justified assumption, because the relaxation time needed to reach this steady state, defined by the ratio of the disc mass to the mass transfer rate can be long; for example, in the case shown in Fig. \ref{lcurve}, it is 50 days. This implies that, if the mass transfer rate varies on timescales of this order or shorter, the outcome of the magnetic gating instability will depend on the history of the mass transfer rate and might explain why the existence of outbursts does not depend only on the mass transfer rate.

\begin{acknowledgements}
We acknowledge with thanks the variable star observations from the AAVSO International Database contributed by observers worldwide and used in this research. JPL was supported in part by a grant form the French Space Agency CNES.
\end{acknowledgements}

\bibliographystyle{aa}
\bibliography{biblio}
\begin{appendix}
\section{Light curves of V1062 Tau and CTCV J2056-3014}
We show here the light curves of V1062 Tau and CTCV J2056-3014 during outbursts.

\begin{figure}
   \centering
   \includegraphics[width=\columnwidth]{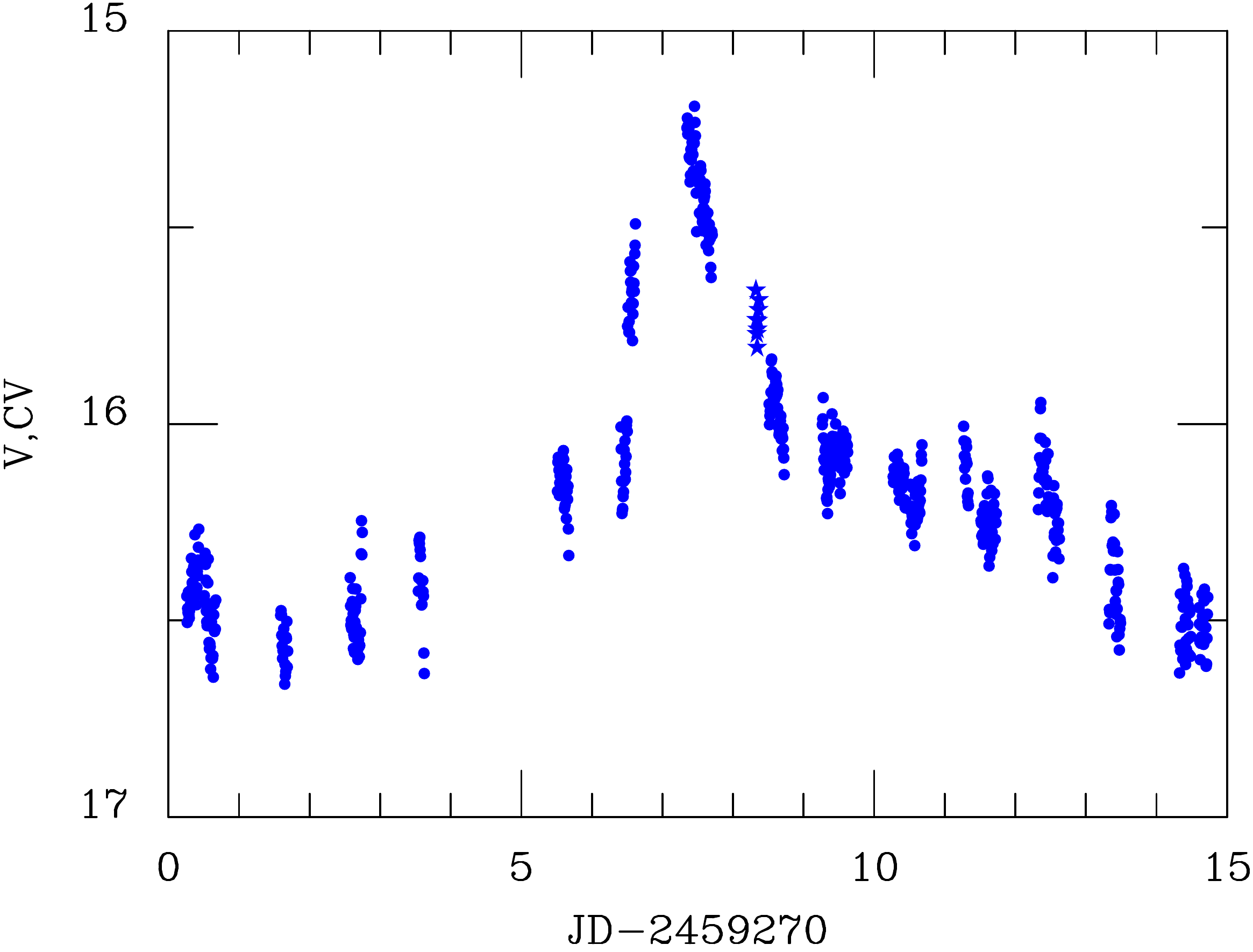}
   \caption{Time profile of the outburst of V1062 Tau (data from AAVSO). The visual $V$ and $CV$ data are shown by dots and stars respectively.}
   \label{v1062}%
   \end{figure}

V1062 Tau has an orbital period of 9.95~h and a spin period of 0.62~min \citep{hbm02}, and was known for having had short (1--2 days) and weak (less than 1 mag.) outbursts \citep{lbo04}; because of a limited sampling, the full outburst duration could not be covered, making it difficult to estimate its characteristics, or even to ascertain that these were real bursts. Figure \ref{v1062} shows the time profile of what appears to be a standard, but faint, outburst. Its amplitude was 1.2 mag and it lasted for about a week. Data in the CV band were obtained by 

\begin{figure}
   \centering
   \includegraphics[width=\columnwidth]{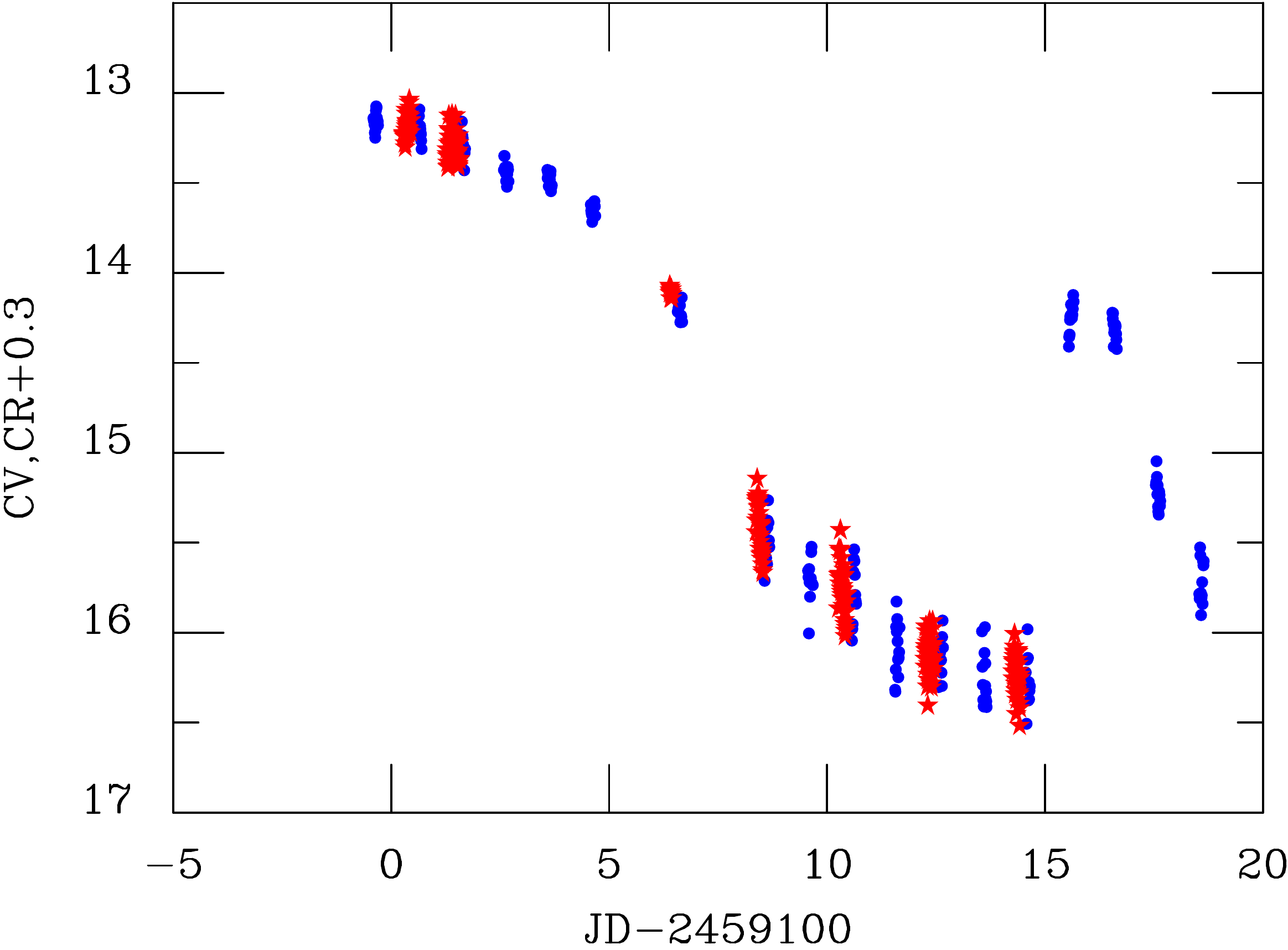}
   \caption{Time profile of the superoutburst of CTCV J2056-3014 (data from AAVSO). The visual $CV$ and red $CR+0.3$ data are shown by blue dots and red stars respectively.}
   \label{ctcv}%
   \end{figure}

\begin{figure}
   \centering
   \includegraphics[width=\columnwidth]{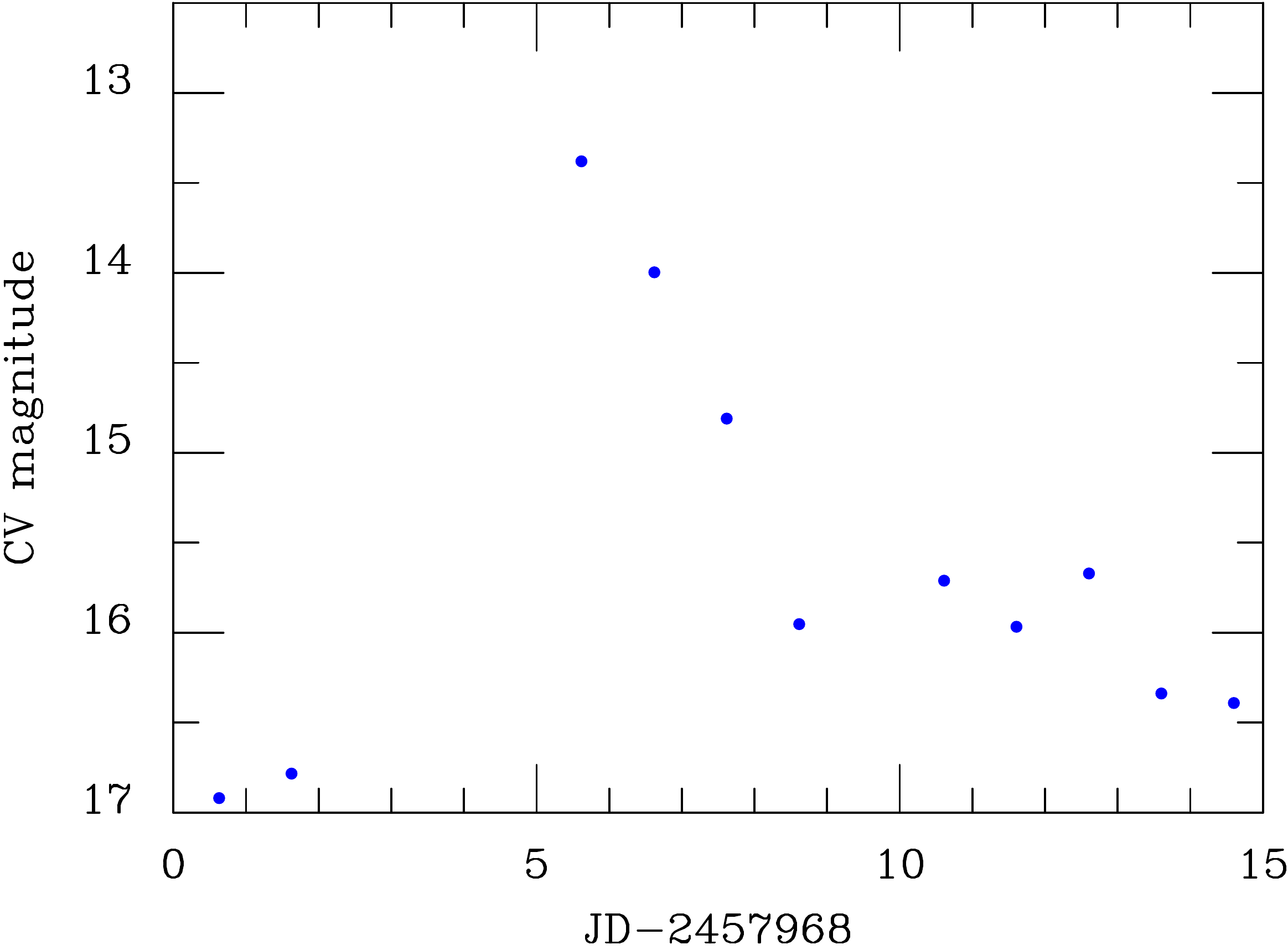}
   \caption{Time profile of the 2017 normal outburst of CTCV J2056-3014 (data from AAVSO).}
   \label{ctcv2}%
   \end{figure}

CTCV J2056-3014 is an intermediate polar with an orbital period of 1.76~h \citep{atd10}. \citet{lbr20} found that the white dwarf is spinning very rapidly, with a spin period of 29.6~s , which means that the magnetic field is low, presumably of order of $6 \times 10^{30}$G~cm$^3$. This source was known to be variable, with strong flickering \citep{b18}, and classified as a dwarf nova in the \citet{rk03} catalogue. The time profile is shown in Fig. \ref{ctcv}. A normal outburst lasting for about five days was also detected in 2017; its maximum was on August 8th. Its time profile is shown in Fig. \ref{ctcv2}. Data in the CV band were obtained by Franz-Josef Hambsch, and data in the CR band by Libert Monard. 

\end{appendix}

\end{document}